\documentstyle[sprocl,epsf]{article}

\def\Journal#1#2#3#4{{#1} {\bf #2}, #3 (#4)}

\def\NPB{{\em Nucl. Phys.} B}
\def\PLB{{\em Phys. Lett.}  B}

\def\PRC{\em Phys. Rev. C}
\def\PR{\em Phys. Rev.}
\def\PRD{{\em Phys. Rev.} D}

\def\be{\begin{equation}}
\def\ee{\end{equation}}
\def\bea{\begin{eqnarray}}
\def\eea{\end{eqnarray}}
\def\MSbar{$\overline{\rm MS}\hbox{ }$}
\def\nospMSbar{$\overline{\rm MS}$}

\begin{document}

\title{REGULARIZATION AND THE POTENTIAL OF EFFECTIVE FIELD THEORY
IN NUCLEON-NUCLEON SCATTERING~\footnote{A more detailed description
of much of the work discussed here was given in Ref.~\cite{Be97B}.}}

\author{D. R. PHILLIPS}

\address{Department of Physics,\\ University of Maryland,\\
College Park, MD 20742-4111, USA\\E-mail: phillips@quark.umd.edu}

\maketitle\abstracts{This paper examines the role that regularization
plays in the definition of the potential used in effective field
theory (EFT) treatments of the nucleon-nucleon interaction. I consider
$NN$ scattering in $S$-wave channels at momenta well below the pion
mass. In these channels (quasi-)bound states are present at energies
well below the scale $m_\pi^2/M$ expected from naturalness
arguments. I ask whether, in the presence of such a shallow bound state,
there is a regularization scheme which leads to an EFT potential that
is both useful and systematic. In general, if a low-lying bound state
is present then cutoff regularization leads to an EFT potential which
is useful but not systematic, and dimensional regularization with
minimal subtraction leads to one which is systematic but not useful.  The
recently-proposed technique of dimensional regularization with
power-law divergence subtraction allows the definition of an EFT
potential which is both useful and systematic. }

\section{The potential of effective field theory}

As described elsewhere in these proceedings~\cite{vK98}, the rise of
effective field theory as a technique in nuclear physics began with
the seminal papers of Weinberg~\cite{We90,We91,We92}. These papers proposed
implementing the EFT program in nuclear physics by applying the
power-counting arguments of chiral perturbation theory to an
$n$-nucleon effective potential rather than directly to the
$n$-nucleon S-matrix. Only $n$-nucleon irreducible graphs should be
included in the $n$-nucleon effective potential. The potential
obtained in this way is then to be inserted into a Lippmann-Schwinger
or Schr\"odinger equation and iterated to all orders. Of course,
unknown coefficients appear in this effective potential, but these can
be fit to experimental data as in ordinary chiral perturbation
theory~\cite{Or96,Ka96,Sc97,Le97,SF98A,SF98B,Pa97}.

Such an EFT treatment of the $NN$ interaction differs in a
fundamental way from conventional EFT applications like $\pi \pi $
scattering in chiral perturbation theory. In both cases operators are
ordered in an effective Lagrangian in the same way. However, in $\pi
\pi $ scattering the operator expansion in the effective Lagrangian
maps to a power series in $k/M$ in the scattering amplitude. It is
straightforward to see that EFT treatments where there is a direct
mapping from the Lagrangian to the S-matrix are
systematic~\cite{We79}. On the other hand, when the mapping is from
the Lagrangian to an effective potential which is subsequently
iterated to all orders, many issues arise which lead one to question
the existence of a systematic power counting in the potential.

In order to discuss some of these issues I consider $NN$ scattering in
the ${}^1S_0$ channel at momentum scales $k \ll m_\pi$.  The EFT at
these scales involves only nucleons since the pion is heavy and may
therefore be ``integrated out''.  The effective Lagrangian thus
consists of contact operators of increasing dimensionality constrained
by spin and isospin. In any $S$-wave channel the operators which
contribute take the form~\cite{Ka96}:
\begin{equation}
{\cal L}=N^\dagger i \partial_t N - N^\dagger \frac{\nabla^2}{2 M} N
- \frac{1}{2} C (N^\dagger N)^2\\ 
-\frac{1}{2} C_2 (N^\dagger \nabla^2 N) (N^\dagger N) + h.c. + \ldots.
\label{eq:lag}
\end{equation}
I do not intend that this EFT should provide a quantitative
description of the $NN$ phase shifts. Instead, I study it because the
scattering amplitudes can be calculated analytically. It therefore
allows the elucidation of issues of principle in EFT for $NN$ scattering.

Such a Lagrangian leads to the following expansion of the potential in
$S$-wave channels:

\begin{equation}
V(p',p)=C + C_2 (p^2 + p'^2) + C_4
(p^4 + p'^4) + C_4' p^2 p'^2 +\dots.
\label{eq:Vexp}
\end{equation}
The prejudice of effective field theory is that the coefficients in
the potential should be ``natural'', i.e. $C_{2n} \sim \frac{1}{m_\pi^{2 +
2n}}$. The potential (\ref{eq:Vexp}) is then to be iterated via the
Lippmann-Schwinger (LS) equation:

\begin{equation}
T(p',p;E)=V(p',p) + M\int \frac{d^3q}{(2 \pi)^3} \, V(p',q) 
\frac{1}{EM- {q^2}+i\epsilon} T(q,p;E).
\label{eq:LSE}
\end{equation}
The hope is that by adopting this procedure one can generate
two-nucleon (quasi-)bound states at the experimentally-observed
energies, which are ``unnaturally'' low, while maintaining ``natural''
coefficients in the potential. In doing this the expansion
(\ref{eq:Vexp}) must be truncated at some finite order in the
quantities $p/m_\pi$ and $p'/m_\pi$. Provided $p,p' \ll m_\pi$ the
neglected terms will be small.

Upon iteration of any truncation of (\ref{eq:Vexp}) ultraviolet
divergences arise. Hence, non-perturbative regularization and
renormalization are required when iterating to all orders using the LS
equation. Divergences arise because in solving the LS equation one
integrates the potential over all momenta. Of course, the expansion
(\ref{eq:Vexp}) is not a truthful representation of the physics of the
$NN$ potential for $p,p' > m_\pi$. Thus, a key question is whether all
divergences can be regularized and renormalized in such a way that the
momentum scales probed inside loops are ultimately well below
$m_\pi$. If this cannot be done then it follows that the expansion
(\ref{eq:Vexp}) will be being used outside its domain of validity, and the
$NN$ potential derived from effective field theory will not be any
more systematic than the many phenomenological $NN$ potentials on the
market.

So, the key question which I address in this paper is whether there is
a regularization technique which leads to an expansion for the $NN$
potential that is both:
\begin{itemize}
\item systematic, i.e. can be truncated at a finite order and is then
used only within the domain of validity of this truncation; and

\item useful, i.e. when put in the regularized Lippmann-Schwinger
equation, provides a reasonable description of the $NN$ phase shifts
in what we would expect to be the domain of validity of the EFT, $k <
m_\pi$.
\end{itemize}

Here this question is addressed using three different regularization
methods. In Section~\ref{sec-cutoff} I discuss the regularization of
the interaction by a momentum-space cutoff. I show how one can
renormalize the coefficients in the potential, and how such an
approach can provide a valid description of the $NN$ scattering data
in the ${}^1S_0$ channel. This description can be progressively
improved by adding more coefficients in the effective
potential. However, in spite of this success, I will show that the
typical momentum inside loops in the Lippmann-Schwinger equation is
such that the potential is being used in a region where to truncate it
at any finite order is not a justified procedure. (Similar conclusions
are discussed elsewhere in this volume~\cite{Bi98} and in
Ref.~\cite{Ri97}.) In Section~\ref{sec-DRMS} I review the failure of
dimensional regularization with minimal subtraction as a candidate for
the role of regulator in the Lippmann-Schwinger equation. In
particular, I will explain why this approach leads to an expansion
which is systematic but not useful. In Section~\ref{sec-DRPDS} the use
of the so-called PDS modification of dimensional regularization in
this problem~\cite{Ka98A,Ka98B,Ka98C} will be discussed. In
particular, this scheme satisfies both of the above criteria, although
the sense in which one is learning something about the underlying $NN$
potential remains somewhat unclear. Finally, in
Section~\ref{sec-conclusions} I will offer some conclusions.

\section{Effective Field Theory with Cutoffs}

\label{sec-cutoff}

In this section I investigate the possibility of an EFT for $NN$
scattering in the presence of a finite cutoff. This physically
intuitive approach has been advocated by Lepage~\cite{Le97,Le90}.  The
idea is that one takes an underlying theory of $NN$ interactions and
introduces a (sharp or smooth) momentum cutoff $\beta$ representing
the scale at which the first new physics becomes important. All loops
now only include momenta $p < \beta$. Of course, one must compensate
for the effects of the neglected modes. However, Lepage argues that
since these modes are highly virtual, one may approximate their
effects by a sequence of local contact interactions. Furthermore, if
the cutoff $\beta$ is placed well below the mass $\Lambda$ of some
exchanged quantum, then, for momenta ${\bf p}$ and ${\bf p}'$ below
the cutoff, the exchange of this quantum:
\begin{equation}
V_\Lambda({\bf p}',{\bf p}) \sim \frac{1}{({\bf p}' - {\bf p})^2 + \Lambda^2};
\end{equation}
may be replaced by a contact interaction, since $p',p < \beta <
\Lambda$.  Therefore the effects coming from exchanges of quanta with
masses well above the cutoff scale $\beta$ may also be approximated by
contact interactions. For the numerical application of these ideas to
the $NN$ problem see~\cite{Or96,Sc97,Le97,SF98A,SF98B,Pa97}.

Now, all that has been said in the previous paragraph still applies if
the cutoff $\beta$ is set below the scale $m_\pi$. Then the only
explicit degrees of freedom in the problem are nucleon modes with
momenta below $\beta$. All higher-momentum nucleon modes {\em and} all
exchanged mesons are integrated out. This cutoff effective field
theory of the $NN$ interaction is of little practical use, but can be
investigated analytically in a way that raises issues of
principle. The effective Lagrangian is that of Eq.~(\ref{eq:lag}).
The effective potential which corresponds to this Lagrangian now
includes theta functions which introduce a sharp cutoff:

\begin{equation}
V(p',p)=[C + C_2 (p^2 + p'^2) + \ldots] \theta(\beta - p)
\theta(\beta - p'),
\label{eq:Veffexp}
\end{equation}
so all integrals
(not just the divergent ones) will be cut off sharply at momentum
$\beta$. After renormalization the coefficients $C$, $C_2$, etc.,
will, of course, depend on the cutoff scale $\beta$, as well as on
physical scales in the problem.

Of course, the expression (\ref{eq:Veffexp}) is an infinite series,
and for practical computation some method of truncating it must be
found.  The fundamental philosophy of cutoff EFT provides a rationale
for this as follows. If we work to any finite order in the effective
potential, cutoff-dependent terms in the scattering amplitude will
appear. These must be in correspondence with neglected higher-order
operators in $V$.  If such terms are progressively added to $V$, one
may remove the cutoff dependence order-by-order. Below we will see
that one can indeed define such a ``systematic'' procedure in this
problem. However, {\it in order that it really make sense to truncate
the effective potential (\ref{eq:Veffexp}) at some finite order it
must be that the operators which are neglected are in some sense
small.}

\subsection{``Second order'' cutoff EFT calculation}
\label{sec-nu2ceft}

I now investigate whether this is indeed the case by looking at the
amplitude and renormalization conditions which arise when one takes the
``second-order'' effective potential:
\begin{equation}
V^{(2)}(p',p)=[C + C_2 (p^2 + p'^2)] \theta(\beta - p)
\theta(\beta - p'),
\end{equation}
and iterates it via the Lippmann-Schwinger equation. Standard
methods~\cite{Ri97,Ph97} lead to the amplitude:
\begin{equation}
\frac{1}{T^{\rm on}(k)}=\frac{(C_2 I_3 -1)^2}{C + C_2^2 I_5 + {k^2} C_2 (2 -
C_2 I_3)} - {\cal I},
\label{eq:secondorderamp}
\end{equation}
where 
\begin{equation}
I_n \equiv -M \int \frac{d^3q}{(2 \pi)^3} q^{n-3},
\label{In}
\end{equation}
and 
\begin{eqnarray}
{\cal I} &\equiv& -M \int \frac{d^3 q}{(2 \pi)^3} \frac{1}{k^2 -
q^2 + i \eta} \label{eq:calI}\\ 
&=& I_1 - \frac{i M k}{4 \pi} + M k^2 {\cal P} \int \frac{dq}{2 \pi^2}
\frac{1}{k^2 - q^2}. \label{eq:calI2}
\end{eqnarray}
In this calculation all integrals are understood to be sharply cutoff
at $q=\beta$, and $k=\sqrt{ME}$ is the on-shell momentum.  The ${\cal
P}$ in Eq.~(\ref{eq:calI2}) indicates a principal value integral.

I now renormalize by demanding that, up to terms of $O(k^4)$,
Eq.~(\ref{eq:secondorderamp}) reproduce the inverse amplitude obtained
when only the first two terms in the effective range expansion are
retained
\begin{equation}
\frac{1}{T^{\rm on}(k)}=-\frac{M}{4 \pi}\left[-\frac{1}{a} + \frac{1}{2} r_e 
k^2 - i k\right].
\label{eq:reamp}
\end{equation}
This yields the following equations for $C$ and $C_2$:
\begin{eqnarray}
\frac{M}{4 \pi a}&=&\frac{(C_2 I_3 -1)^2}{C + C_2^2 I_5} - {I_1};
\label{eq:Cbeta}\\
\frac{M r_e}{8 \pi}&=&\left(\frac{M}{4 \pi a} + I_1\right)^2
\frac{C_2 (2 - C_2 I_3)}{(C_2 I_3 - 1)^2} + \frac{M}{2 \pi^2 \beta},
\label{eq:C2beta}
\end{eqnarray}
where the last term in Eq.~(\ref{eq:C2beta}) arises because the
presence of the cutoff generates additional energy dependence when the
integral ${\cal I}$ is evaluated.
 
\begin{figure}[h,t,b]
   \vspace{0.5cm} \epsfysize=8 cm \centerline{\epsffile{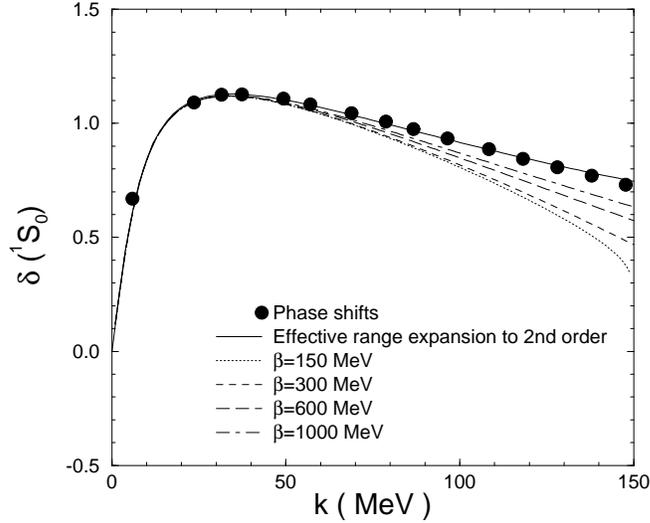}}
   \centerline{\parbox{11cm} {\caption{\label{fig-plot} Phase shifts
   in the $NN$ ${}^1S_0$ channel predicted for different
   values of the cutoff $\beta$.  The dots are taken from the
   Nijmegen phase shift analysis of Ref.~\protect\cite{St93}. The
   solid line is the effective range expansion to second order, which
   is reproduced when $\beta \rightarrow \infty$. The other lines
   represent the different choices for the cutoff $\beta$ indicated in
   the legend.}}}
\end{figure}

Once $\beta$ is fixed these equations can be solved for $C$ and $C_2$.
Of course, as $\beta$ is varied the $C$ and $C_2$ that satisfy
Eqs.~(\ref{eq:Cbeta}) and (\ref{eq:C2beta}) will change
significantly. However, because one is fitting to low-energy
scattering data different values of $\beta$ will not lead to any
fundamental differences in the low-energy T-matrix. Since $C$ and
$C_2$ are fit to the first two terms in the effective-range expansion
sensitivity to the cutoff appears in the on-shell inverse amplitude at
order $(k/\beta)^4$. The results of this procedure for $NN$ phase
shifts in the ${}^1S_0$ channel are shown in Fig.~\ref{fig-plot}, for
a range of cutoff values, from $\beta=150$ MeV to $\beta=\infty$. Also
shown are the $NN$ phase shifts in the Nijmegen group's phase shift
analysis~\cite{St93}, and the amplitude (\ref{eq:reamp}). Note that
this amplitude yields an excellent fit to the experimental data up to
momenta $k$ of order $m_\pi$. The amplitudes obtained in the cutoff
EFT also do a reasonable job of describing the data, especially as the
cutoff is increased. These results, and the similar in spirit though
much more thorough studies of Refs.~\cite{Le97,SF98A,SF98B}, show that
the potential in cutoff effective field theory satisfies the second of
our two conditions: it is useful.

Furthermore, working to higher order in the potential will
progressively improve these fits. If I constructed the ``next-order''
effective potential, $V^{(4)}$ and refitted the (four) coefficients
appearing in it to the first four terms in the effective range
expansion, then, by construction, sensitivity to the cutoff will appear
in $k \cot \delta$ at $O((k/\beta)^8)$. In this sense the introduction
of higher-dimension operators does allow for the ``systematic''
removal of cutoff dependence in the amplitude. This is the
systematicity seen in the work of Scaldeferri {\it et
al.}~\cite{Sc97}, Lepage~\cite{Le97}, and Steele and
Furnstahl~\cite{SF98A,SF98B}. Nevertheless, this does not yield a {\it
systematic EFT potential} in the sense I have defined it here.  After
all, {\it any} parameterization of a potential which is rich enough to
successively fit the terms in the effective-range expansion will be
similarly improved as one adds parameters to it. The question I
address here is whether one can use the power counting to argue {\it a
priori} that certain contributions to the potential of
Eq.~(\ref{eq:Veffexp}) will be systematically small and hence can be
neglected at some specified level of accuracy.

\subsection {Power counting in this effective field theory}

To answer this question I must examine the values of $C$ and $C_2$
which are required to solve the equations above.  Assuming that the
cutoff obeys $1/a \ll \beta$ it follows that the second of these two
equations becomes
\begin{equation}
\frac{M}{8 \pi}\left(r_e - \frac{4}{\pi \beta}\right) \approx \frac{I_1^2}{I_3}
\frac{\overline{C}_2 (2 - \overline{C}_2)}{(\overline{C}_2 - 1)^2},
\label{eq:recondn}
\end{equation}
where $\overline{C}_2 \equiv C_2 I_3$. This leads to a quadratic
equation for $\overline{C}_2$, which for values of $\beta$ up to some
$\beta_{\rm max}=\frac{16}{\pi r_e}$ has real solutions.  For $\beta >
\beta_{\rm max}$ the renormalization condition for $C_2$ has no real
solution if $r_e >0$. This is related to the fact that a Hermitian
potential of range $R$ which is to be used in the Schr\"odinger
equation can only yield an effective range $r_e$ consistent
with~\cite{Wi55,PC97,Fe95}:
\begin{equation}
r_e \leq 2\left(R - \frac{R^2}{a} + \frac{R^3}{a^2}\right).
\label{eq:WB}
\end{equation}
As I take $\beta \rightarrow \infty$ I force the range of our EFT
potential to zero. So, if $r_e > 0$ is to be obtained, the potential
must become non-Hermitian. However, Fig.~\ref{fig-plot} shows that one
still obtains perfectly sensible phase shifts even if this is the
case. Another way to evade the bound (\ref{eq:WB}) and so produce
positive effective ranges with a zero-range force is via the
introduction of a dibaryon field~\cite{Ka97,Bd97,vK97}. This idea is
discussed further in other contributions to these
proceedings~\cite{Co98,vK98B,Bd98}.

As for the scaling of the coefficients $C$ and $C_2$,
Eqs.~(\ref{eq:recondn}) and (\ref{eq:Cbeta}) lead to

\begin{equation}
C_2 \sim \frac{1}{M \beta^3} \Rightarrow C \sim \frac{1}{M \beta}.
\label{eq:CC2beh}
\end{equation}

Note that this behavior does not arise if $a$ is natural, i.e. of
order $1/m_\pi$, and $\beta$ is chosen to be less than $m_\pi$; then
the leading order behavior of the coefficients $C$ and $C_2$ is very
different. In fact,
\begin{equation}
C \sim \frac{1}{M m_\pi}; \qquad C_2 \sim \frac{1}{M m_\pi^2 \beta}.
\end{equation}
In this case all loop effects coming from $C_2$ are suppressed by a
factor of at least $(\beta/m_\pi)^2$. Indeed, all loop effects are
suppressed by a factor of at least $\beta/m_\pi$.  Therefore, if
$\beta < m_\pi$ a non-perturbative calculation is not necessary. In
other words, if the experimental parameters are natural then cutoff
field theory with $\beta < m_\pi$ gives a perturbative EFT in which
loop graphs are consistently suppressed~\cite{Ma95}. However, if a
perturbative calculation is performed then the regularization scheme
chosen becomes immaterial, as the short-distance physics may be
renormalized away. 

On the other hand, for unnaturally long scattering lengths
Eq.~(\ref{eq:CC2beh}) shows that:
\begin{equation}
\frac{C_2}{C} \sim \frac{1}{\beta^2}.
\label{eq:ratio}
\end{equation}
Consequently the condition for us to be able to truncate the expansion
of the potential at zeroth order would be $\hat{p}^2 \ll \beta^2$.  It
is easy to give a heuristic justification of why the behavior
(\ref{eq:ratio}) arises in a non-perturbative cutoff EFT calculation,
and why to expect similar behavior to all orders in the effective
potential. After all, the choice of a theta function to regulate the
momentum-space integrals as in Eq.~(\ref{eq:Veffexp}) is entirely
arbitrary. All that has been said above could be reformulated with a
smooth cutoff. This would result in an effective potential of the form
\begin{equation}
V(p',p)=[\tilde{C} + \tilde{C_2} (p^2 + p'^2) + \ldots]
g(p^2/\beta^2,p'^2/\beta^2)
\label{eq:Veffexp2}
\end{equation}
where $g(x,y)$ obeys $g(0,0)=1$, $g(x,y)=g(y,x)$ and $g(x,y)
\rightarrow 0$ faster than any power of $x$ as $x \rightarrow \infty$
with $y$ held fixed. In a non-perturbative calculation the effective
potential should be essentially unaltered by this change in the form
of the cutoff. However, this necessarily means that the ratios
$\tilde{C_{2n}}/\tilde{C}$ differ from those $C_{2n}/C$ by terms of order
$1/\beta^{2n}$.  Therefore for a generic cutoff function $g$ the 
ratio $C_{2n}/C$ must be of order $1/\beta^{2n}$.

Now, if the ratio $C_{2n}/C$ goes like $1/\beta^{2n}$, then 
in order to justify a systematic truncation of the
effective potential we require $\hat{p}^{2n} \ll \beta^{2n}$. However,
the effective potential is to be used in a momentum regime which
extends up to $\beta$, and at the upper end of this momentum regime it
is clear that all terms in the expansion for $V$ are
equally important.

Of course, if internal loops were dominated by the external momentum,
$k$, and so $\hat{p} \approx k$, then this behavior of the
coefficients would not be cause for concern, since $k \ll \beta$ can
be maintained. However, virtual momenta up to $\beta$ flow through all
internal loops. Therefore I believe it makes sense to consider a
quantum average in testing to see if $\hat{p}^{2n} \ll \beta^{2n}$,
since such an average is sensitive to these virtual effects.

All arguments about the size of operators in the effective action
would apply equally well if there was a low-energy bound state in the
channel under consideration. So, let us evaluate quantum averages of
the operator $\hat{p}^{2n}$ using the bound-state wave function
obtained from the zeroth-order EFT potential.  The zeroth-order
potential yields a wave function for the bound state of energy $E=-B$,
\begin{equation}
\psi^{(0)}(p)={\cal N} \frac{M}{MB + p^2} \theta(\beta-p),
\end{equation}
where ${\cal N}$ is some normalization constant. For $MB \ll \beta^2$
this gives
\begin{equation}
\frac{\langle \hat{p}^{2n} \rangle}{\beta^{2n}} \equiv \frac{\langle
\psi^{(0)}|\hat{p}^{2n}|\psi^{(0)} \rangle}{\beta^{2n}} = \frac{4}{(2n-1) \pi}
\frac{\sqrt{MB}}{\beta}, \qquad n=1,2,\ldots.
\label{eq:avgs}
\end{equation}
Thus, $\hat{p}^{2n} \ll \beta^{2n}$ is apparently satisfied. However,
Eq.~(\ref{eq:avgs}) shows that if $\langle V \rangle$ is calculated
with the wave function $\psi^{(0)}$ there is no reason to truncate the
expansion at any finite order, since all terms beyond zeroth order
contribute with equal strength to the quantum average.

If there was systematic power counting for the $NN$ potential in
cutoff field theory then the contribution of these ``higher-order''
terms in the potential should get systematically smaller as the
``order'' is increased.  However, it is clear that this does not
happen---rather, all terms beyond zeroth order contribute to the
potential at the same order.  Therefore one cannot justify a
truncation of Eqs.~(\ref{eq:Veffexp}) and (\ref{eq:Veffexp2}) at some
finite order in $p$ and $p'$. Such a truncation may result in a good
fit to the experimental data for on-shell momenta $k \ll \beta$, but
it is not based on a systematic expansion of the $NN$ potential in
powers of momentum.

\section{Dimensional Regularization with Minimal Subtraction}

\label{sec-DRMS}

In ordinary perturbative EFT calculations the existence of a
consistent power-counting scheme in the S-matrix relies on removing
the short-distance physics that arises in loop graphs via the
renormalization procedure~\cite{We79}. Since the short-distance
physics is removed in a manner insensitive to choice of regularization
scheme it is economical to use dimensional regularization (DR) to
regularize and renormalize, because DR respects chiral and gauge
symmetries. When considering the relevance of EFT methods in nuclear
physics one might choose to extrapolate intuition gained from
perturbation theory and regulate the divergent loops in our
Lippmann-Schwinger equation using DR. It is important to realize that
the use of DR implicitly {\it assumes} that the short-distance physics
buried in loop graphs {\it does not} contribute to low-energy physics.

In this section I compare dimensional regularization with minimal
subtraction (DR with \nospMSbar), as implemented by Kaplan {\it et al.} in
their 1996 paper~\cite{Ka96}, and cutoff schemes.  There are two
fundamental points I wish to make
\begin{itemize}
\item The conclusions which one reaches about both the underlying
potential and the usefulness of the effective field theory are
different in the two different regularization schemes.  

\item When low-lying (quasi-)bound states are present, a necessary
(but not sufficient) condition for a workable EFT treatment of $NN$
scattering is that some short-distance effects from loops contribute
to the physical scattering amplitude.
\end{itemize}

The expression (\ref{eq:secondorderamp}) displayed above is in fact
true in any regularization scheme. In regularization schemes other
than the cutoff method the integrals will be defined in different
ways. We now use DR with \MSbar to regulate the integrals $I_1$, $I_3$,
and $I_5$ which appear in Eq.~(\ref{eq:secondorderamp}).
This is a convenient way to implement an idea which is
central to the success of perturbative EFT: the power-law divergent
pieces of integrals over internal loop momenta should not affect the
final physical scattering amplitude. In DR with \MSbar all power-law
divergences vanish, therefore $I_1=I_3=I_5=0$.  Consequently, the
on-shell amplitude takes the form~\cite{Ka96}

\begin{equation}
\frac{1}{T^{\rm on}_{\overline{\rm
MS}}(k)}=\frac{1}{C_{}^{\overline{\rm MS}}+ 2 C_2^{\overline{\rm MS}} k^2} +
\frac{i M k}{4 \pi}.
\label{eq:TDRwithMS}
\end{equation}
Renormalizing so as to reproduce the effective-range expansion to
second order leads to the values~\cite{Ka96}
\begin{equation}
C^{\overline{\rm MS}}=\frac{4 \pi a}{M}; 
\qquad C_2^{\overline{\rm MS}}=\frac{\pi a^2 r_e}{M}.
\label{eq:DRMScoeffts}
\end{equation}

Hence, in this instance we conclude that the expansion of the EFT
potential has a domain of validity $k^2 \ll 2/(a r_e)$, as that is the
point at which the second term in the expansion of the EFT potential
becomes as large as the first. This is a natural region if both $a$
and $r_e$ are natural, as then the EFT is valid for $k^2 \ll
\Lambda^2$. However, if the scattering length is unnaturally large the
momentum domain over which the EFT obtained when DR with \MSbar is
used is small. In fact, as noted by Kaplan {\it et al.}, and shown in
Fig.~\ref{fig-PDSplot}, Eq.~(\ref{eq:TDRwithMS}) reproduces the data
in the ${}^1S_0$ extremely poorly, since it only agrees with the
phenomenologically efficacious amplitude (\ref{eq:reamp}) at very
small $k$. In order to restore the agreement higher-dimensional
operators, all containing powers of the low-energy scale $1/a$, would
have to be added to the theory~\cite{Ka96,vK97,LM97}. Thus, in
contrast to the results obtained with cutoff regularization, where we
concluded that the effective theory was valid for $k^2 \ll \beta^2
\leq m_\pi^2$, here we conclude that the domain of validity of the EFT
is $k^2 \ll 1/(a r_e)$. So, different conclusions about the
effectiveness of the EFT are reached when different regularization
schemes are chosen.

In fact, since there are no logarithmic divergences in this problem
and the loop graphs have no finite real part, it is straightforward to
show that DR with \MSbar gives an amplitude in which only the absorptive
parts of the loop graphs are retained~\cite{Be97B,Le97}. In other
words, using DR with \MSbar is equivalent to making a power-series
expansion in the momentum $k$ for the on-shell K-matrix, and then
unitarizing the result.

Note also that from Eq.~(\ref{eq:DRMScoeffts}) we would infer that the
$NN$ EFT potential was real, in sharp distinction to the conclusions
implied by the Wigner bound. However, this distinction occurs because
$C$ and $C_2$ do not represent coefficients in an expansion of a
quantum mechanical potential. Indeed, as explained above, they really
represent coefficients in an expansion of the on-shell K-matrix.
Thus, the connection of the amplitude $C + C_2 (p^2 + p'^2)$ to
the underlying dynamics is less transparent if DR with \MSbar is used than
in the cutoff approach of Section~\ref{sec-cutoff}.

The use of DR with \MSbar does lead to an effective field theory which
is systematic. Since only the on-shell part of internal loops is
retained the EFT ``potential'' is used at all times only within the
domain where the expansion is valid, provided that on-shell momenta $k
\ll \frac{2}{a r_e}$ are considered.  However, as discussed above,
this is a small energy domain, and so the resulting EFT is not
particularly useful. This occurs because when there are low-lying
bound states the low-energy scale $1/a$ still sets the scale of the
coefficients in the ``potential'' of the DR with \MSbar
calculation. This was the problem which iterating the ``potential''
via the LS equation was supposed to avoid.

However, this failure of dimensional regularization with minimal
subtraction is not particularly surprising when one considers that in
quantum mechanics an unnaturally large scattering length can occur via
the cancelation between a natural ``range'' and a natural
``strength'' of a potential~\cite{Le97,Co97}.  The implementation of
this general cancelation between range and strength must be an
element of any useful EFT description of $NN$ scattering.  DR with \MSbar
discards all short-distance physics that comes from loops. This
follows necessarily from its being a scale-independent regularization
scheme. In general, information about the range of the potential
enters through these power-law divergences, and so DR with \nospMSbar's
neglect of all power-law divergences means it does not retain this
information on the range of the interaction. Hence, in DR with \MSbar the
coefficients in the Lagrangian are forced to take on unnatural sizes,
and although the resulting effective field theory is systematic, it is
not useful.

\section {Dimensional regularization with power-law divergence subtraction}

\label{sec-DRPDS}

This raises the question of whether we can modify the minimal
subtraction scheme in a way that avoids these problems. Recently
Kaplan {\it et al.}~\cite{Ka98A,Ka98B} have proposed a different
subtraction scheme, power-law divergence subtraction (PDS), which
attempts to do precisely this.  This scheme is described more fully
elsewhere in these proceedings~\cite{Ka98C}. Its success is based on
its modification of the usual DR prescription of ignoring all
power-law divergences. In PDS a term corresponding to a linear
divergence is included in the definition of the divergent integral
${\cal I}$, where
\begin{equation}
{\cal I}(k)=\left(\frac{\mu}{2}\right)^{4 - D} M \int
\frac{d^{(D-1)}p}{(2 \pi)^{D-1}} \frac{1}{k^2 - p^2 + i \eta},
\end{equation}
is the integral of Eq.~(\ref{eq:calI}) evaluated in $D$ space-time
dimensions, and $\mu$ is an arbitrary scale.  This is achieved by
adding a piece to ${\cal I}$ which cancels its logarithmic divergence
in $D=3$. When the resulting subtracted integral is continued back to
$D=4$ we get:
\begin{equation}
{\cal I}^{\rm PDS}=-\frac{M}{4 \pi}(i k + \mu).
\label{eq:calIPDS}
\end{equation}
In terms of the integrals $I_1$, $I_3$, and $I_5$ discussed above we
still have $I_3=I_5=0$, but now $I_1=-\frac{M \mu}{4 \pi}$.  The
additional piece added to cancel the logarithmic divergence in $D=3$
corresponds to terms linear in $\beta$ in the cutoff EFT approach
discussed in Section~\ref{sec-cutoff}. Note that PDS reduces to the result
obtained in the \MSbar scheme if $\mu=0$.

From Eq.~(\ref{eq:secondorderamp}) it is trivial to see that when we
solve the effective field theory by taking the second-order tree-level
amplitude, iterating it via the Lippmann-Schwinger equation, and using
DR with PDS to regulate the loops, we get an amplitude
\footnote{Note that in using PDS in this way I am not following the
approach of its inventors. They use the derived scaling of the
coefficients $C^{\rm PDS}(\mu)$ and $C_2^{\rm PDS}(\mu)$ to infer that
all operators beyond the lowest-order one $C (N^\dagger N)^2$ should
be treated perturbatively.  Here I iterate all operators in the
tree-level amplitude to all orders. If the higher-dimensional
operators are truly suppressed then the difference between this
approach and that advocated by Kaplan {\it et al.} should be small.}
\begin{equation}
\frac{1}{T^{\rm on}(k)}=\frac{1}{C^{\rm PDS}(\mu) + 2 C_2^{\rm PDS}(\mu)
k^2} + \frac{M \mu}{4 \pi} + \frac{i M k}{4 \pi}.
\label{eq:TPDS}
\end{equation}
(A similar amplitude has recently been written down
independently~\cite{Ge98}.) In the language used in the previous
section the amplitude (\ref{eq:TPDS}) may be obtained from that found
in the \MSbar scheme by the addition of a constant dispersive part to all
loops.

\begin{figure}[h,t,b]
   \vspace{0.5cm} \epsfysize=8 cm \centerline{\epsffile{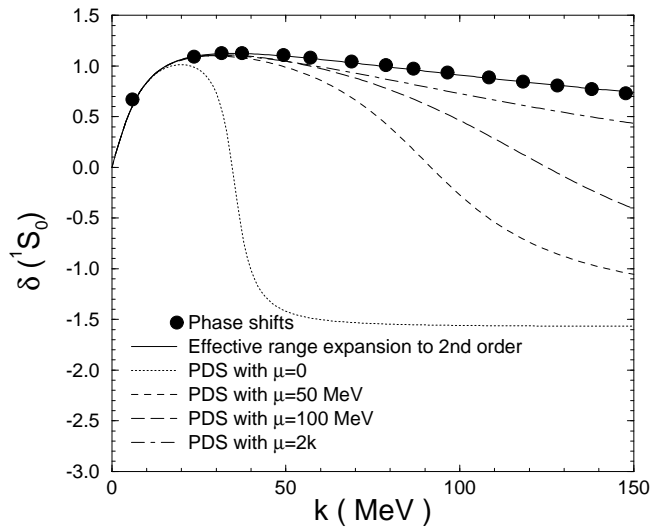}}
   \centerline{\parbox{11cm} {\caption{\label{fig-PDSplot} Phase
   shifts in the $NN$ ${}^1S_0$ channel for different choices of the
   scale $\mu$ of PDS.  The dots are taken from the Nijmegen phase
   shift analysis of Ref.~\protect\cite{St93}. The solid line is the
   effective range expansion to second order, which is reproduced when
   $\mu \rightarrow \infty$. The other lines represent the different
   choices for the parameter $\mu$ indicated in the legend. Note that
   the choice $\mu=0$ yields the \MSbar result.}}}
\end{figure}

As is discussed in Refs.~\cite{Ka98A,Ka98B,Ka98C} it is now a simple
matter to show that, with a suitable choice of $\mu$, this approach is
both systematic and useful. The coefficients $C(\mu)$ and $C_2(\mu)$
may be adjusted to fit the scattering length and effective range. This
leads to~\cite{Ka98A,Ka98B}
\begin{equation}
C_{}^{\rm PDS}(\mu)=\frac{4 \pi}{M}\frac{1}{-\mu + 1/a}; \qquad
C_2^{\rm PDS}(\mu)=\frac{4 \pi}{M} \left(\frac{1}{-\mu+1/a}\right)^2
\frac{r_e}{4}.
\end{equation}
The PDS amplitude with various choices of $\mu$ is compared with the
effective range expansion in Fig.~\ref{fig-PDSplot}. Note that a good
result is obtained when $\mu$ is chosen to be of order, but larger,
than $k$. Thus, with a suitably large choice for $\mu$, PDS gives 
a useful EFT for $NN$ scattering in the ${}^1S_0$ channel.

The EFT thus obtained is also systematic, since in a theory with an
unnatural scattering length and a natural effective range the scaling
of the coefficients for $\Lambda \gg \mu \gg 1/a$
is~\cite{Ka98A,Ka98B}
\begin{equation}
C_{2n} \sim \frac{1}{M \mu^{n+1} \Lambda^n}.
\end{equation}
Now we see that there will be a valid reason to truncate the expansion
for $V$:
\begin{equation}
V=\sum_{n=0}^\infty C_{2n} {\hat p}^{2n},
\end{equation}
since terms beyond zeroth order will be suppressed by powers of the
on-shell momentum over $\Lambda$. 

Thus, in PDS Kaplan {\it et al.} have proposed a definite solution to
the difficulties which have beset the effective field theory in the
$NN$ interaction. They have modified the usual DR with \MSbar prescription
just enough to implement the cancelation between range and strength
which allows quantum mechanical potentials with natural strengths and
sizes to lead to unnaturally shallow bound states.  In doing so they
have generated an expansion for the $NN$ interaction which is both
useful and systematic. However, it should be noted that the
interaction thus obtained is not really a usual quantum mechanical
``potential'', since PDS, by construction, does not include all of the
effects of virtual momenta inside the loops.

I will conclude this section by raising two questions about PDS which
intrigue me:
\begin{enumerate}
\item Are other modifications of DR with \MSbar possible, and what do
they do to the power counting? In particular, why doesn't the
introduction of a term which mimics the effects of terms linear in a
cutoff parameter lead to the scaling of the coefficients found
in Section~\ref{sec-cutoff}? If PDS was completely equivalent to a
cutoff theory, and $\mu$ played the role of the cutoff then we would
expect the scaling
\begin{equation}
C_{2n} \sim \frac{1}{M \mu^{2n+1}}.
\label{eq:C2nmu}
\end{equation}
If such scaling behavior were found we would be forced to conclude
that power-counting had broken down again, and so PDS's avoiding this
behavior is a key ingredient to its success.  Consequently, it would
be interesting to know why PDS does not lead to the behavior
(\ref{eq:C2nmu}).

\item The demand that all power-law divergences vanish in DR
with \MSbar can be traced to DR's requirement that loop integrals such as
\begin{equation}
\int d^dq \frac{1}{q^n}
\label{eq:divgt}
\end{equation}
should be scale invariant. This is an extremely useful property of DR,
since it makes the preservation of chiral and gauge symmetries much
easier than it is in a scale-dependent regularization such as
Pauli-Villars. However, PDS also leads to a scale-dependent definition
for integrals such as (\ref{eq:divgt}), since they
acquire a dependence on $\mu$.  It remains to be seen how this
presence of an artificial scale in such integrals affects gauge and
chiral invariance.
\end{enumerate}

\section{Conclusion}

\label{sec-conclusions}

In summary, neither of the two obvious regularization methods, DR with
\nospMSbar, or a cutoff, lead to an expansion for the $NN$ interaction
which is both systematic and useful. The expansion obtained using a
cutoff gives a reasonable fit to the phase shift data, but from it we
learn nothing about the underlying short-range potential, since
operators neglected in that potential are generally not smaller than
those included. By contrast, we come to different conclusions about
both the domain of validity of the effective theory and the
systematicity of the approach when DR with \MSbar is used.  DR with
\MSbar fails to give a useful description of the $NN$ amplitude,
because it does not implement the cancelation between ``range'' and
``strength'' which allows potentials with hadronic-scale ranges and
depths to produce bound states with nuclear energies . By modifying DR
with \MSbar to include part of the effects of the range of the
potential, namely the linear divergences, Kaplan {\it et al.} have
constructed a {\it useful and systematic} approach to the effective
field theory of the $NN$ interaction. However, it is not clear how the
resulting coefficients $C_{2n}$ which are defined in the PDS scheme
would be related to any underlying $NN$ potential.

So, it seems that regardless of the regularization scheme used one
actually learns little about the short-range $NN$ potential. Indeed
the meaning of the operator $C + C_2(p^2 + p'^2)$ is completely
different in the three different regularization schemes
considered. This is not necessarily a problem. After all, the quantum
mechanical potential is not an observable. However, if we wish to
systematically use chiral expansions of the sort proposed by Weinberg
in order to describe the physics of the $NN$ potential in a controlled
way we need to be reassured that the momenta for which we are
expanding the potential are small. This is not obviously true inside
loops in the Lippmann-Schwinger equation, and that casts doubt on the
validity of such a program.

Salvation for Weinberg's arguments occurs because nuclear wave
functions are typically dominated by momenta of scale $m_\pi$ or
below.  Once wave functions with those scales are generated, whether
by fair means or foul, Weinberg's power-counting may enable the
inclusion of the physics of pion exchanges in a way that respects
chiral symmetry and is systematically
improvable~\cite{CP98}. Therefore, while the potential of effective
field theory is not observable, the potential of effective field
theory in nuclear physics remains large and attractive.

\section*{Acknowledgments}

I am very grateful to those with whom the work described in this paper
was done, Silas Beane and Tom Cohen, for wrestling with these issues
with me.  I also thank M.~C.~Birse, D.~B.~Kaplan, W.~D.~Linch,
J.~A.~McGovern, M.~J.~Savage, U.~van Kolck, and S.~J.~Wallace for
discussions on this intriguing topic. It is a pleasure to thank the
organizers of the workshop, and the bodies funding it, Caltech and the
Institute for Nuclear Theory, for a very useful and most enjoyable two
days. This research was supported in part by the U. S. Department of
Energy, Nuclear Physics Division (grant DE-FG02-93ER-40762).

\end{document}